\documentclass[twocolumn,prl,showpacs,superscriptaddress]{revtex4}
\usepackage{graphicx}
\usepackage{subfigure}
\usepackage{amsmath}
\begin{document}

\title{Spatial Extent of Random Laser Modes}
\date{\today}
\author{Karen L. van der Molen}
\email{k.l.vandermolen@utwente.nl} \affiliation{Complex Photonic
Systems, MESA$^+$ Institute for Nanotechnology and Department of
Science and Technology\\ University of Twente, P.O. Box 217, 7500 AE
Enschede, The Netherlands.}
\author{R. Willem Tjerkstra}
\affiliation{Complex Photonic Systems, MESA$^+$ Institute for
Nanotechnology and Department of Science and Technology\\ University
of Twente, P.O. Box 217, 7500 AE Enschede, The Netherlands.}
\author{Allard P. Mosk}
\affiliation{Complex Photonic Systems, MESA$^+$ Institute for
Nanotechnology and Department of Science and Technology\\ University
of Twente, P.O. Box 217, 7500 AE Enschede, The Netherlands.}
\author{Ad Lagendijk}
\affiliation{FOM Institute for Atomic and Molecular Physics
(AMOLF), Kruislaan 407, 1098 SJ Amsterdam, The Netherlands}

\begin{abstract}
We have experimentally studied the distribution of the spatial
extent of modes, and the crossover from essentially single-mode to
distinctly multi-mode behavior inside a porous gallium phosphide
random laser. This system serves as a paragon for random lasers due
to its exemplary high index contrast. In the multi-mode regime we
observed mode competition. We have measured the distribution of
spectral mode spacings in our emission spectra, and found level
repulsion that is well described by the Gaussian orthogonal ensemble
of random-matrix theory.
\end{abstract}


\maketitle Random lasers, in which gain is combined with multiple
scattering of light \cite{Letokhov1968,Lawandy1994}, are the subject
of active research \cite{Vutha2006,Sharma2006,Noginov2006,Rose2005}.
The discovery of intriguingly narrow spectral emission peaks
(spikes) in certain random lasers by Cao \textit{et al.}
\cite{Cao1998} has given an enormous boost to this field of research
\cite{Cao2005,Yu2004,Frolov1999,Polson2005,Milner2005,Noginov2004,Mujumdar2004}.
In the past few years a debate on the origin of these spikes has
started
\cite{Mujumdar2004,Sebbah2002,Cao2002,Apalkov2002,Markushev2005,Deych2005,Angelani2006,Pinheiro2006,Stone2006}.
On one hand Cao \textit{et al.} \cite{Cao2002}, as well as Sebbah
and Vanneste \cite{Sebbah2002} and Apalkov \textit{et al.}
\cite{Apalkov2002}, attribute the spikes to local cavities (local
modes, LM) for light, which are formed by multiple scattering. On
the other hand Mujumdar \textit{et al.} \cite{Mujumdar2004}, as well
as Pinheiro and Sampaio \cite{Pinheiro2006}, attribute the spikes to
single spontaneous emission events that, by chance, follow very long
light paths (open modes, OM) in the sample and hence pick up a very
large gain. The LM and the OM model we cite represent divergent
answers to the question what the spatial extent is of the modes
responsible for the spikes.

The spatial extent of the modes is a crucial factor for the
fundamental behavior of a random laser. If the spatial extent of the
modes is small, and the modes do not spatially overlap, the laser
effectively consists of a collection of single-mode lasers. In
contrast, spatially overlapping modes inside the random laser lead
to distinctly multi-mode behavior, such as mode competition
\cite{Ambartsumyan1970}.

In this Letter, we present the first systematic study of the spatial
extent of the modes of a random laser, and show the crossover from
essentially single-mode to multi-mode behavior in a new type of
random laser: single crystal porous gallium phosphide
\cite{Schuurmans1999}, filled with liquid dye solution. Among other
conclusions, our results indicate that the LM model describes the
physics of our random laser better than the OM model, but the LM
model needs more sophistication.

Our random laser consists of porous gallium phosphide obtained by
anodic etching (porosity, 45\% air; thickness, $> 46 \mu$m), of
which the pores are filled with a 10 mmol/l solution of rhodamine
640 perchlorate in methanol (pump absorption length, 22~$\mu$m;
minimal gain length, 12~$\mu$m \cite{Baumler1992}). This random
laser combines three characteristics which sets it apart from other
random lasers. Firstly, the position of the scatterers is rigidly
fixed. Secondly, the efficient laser dye rhodamine 640 ensures high
gain. Finally, the contrast of the refractive indices in the sample
is the highest ever reported for random lasers: n = 1.33 for
methanol and 3.4 for gallium phosphide. The method of fabrication of
the porous gallium phosphide is slightly modified from the method
described earlier \cite{Erne1996,Tjerkstra2002}. We used n-GaP with
a dopant density of 2$\cdot$10$^{17}$/cm$^{2}$, obtained from
Marketech. To make porous GaP without a non-porous top layer, the
GaP is anodically etched in 0.5 M H$_{2}$SO$_{4}$. In the first
step, the potential is set to 22.5 V until 5 C/cm$^{2}$ has passed
through the sample. In the next step, the potential is switched to
30 V for 10 minutes. In the third step, the potential is switched
back to 22.5 V. The GaP is etched until 60 C/cm$^{2}$ has passed
through the sample. After finishing the etching process, the first
layer can easily be removed, leaving porous GaP.

The transport mean free path $\ell$ of our samples filled with pure
methanol is measured with an enhanced-backscatter-cone experiment
\cite{Albada1985} and is 0.6 $\pm$ 0.3~$\mu$m at $\lambda = 633$~nm
($k_0 \ell = 6.4$~$\pm$~2.8, with $k_0$ the vacuum wavenumber,
n$_{\rm{eff}}$ = 2.1 $\pm$ 0.4). The sample holder has a sapphire
window with a thickness of 3 mm. The random laser sample is lightly
pushed onto the window with a spring. A detailed calculation shows
that the emitted laser-light is generated inside the pores of the
gallium phosphide, and cannot be generated by the dye between the
window and the sample. The sample holder can be translated in one
direction (parallel to the surface of the sample) with a total
displacement of 12~mm by means of a stepper motor. The pump source
is a tunable optical parametric oscillator (Coherent Infinity
40-100/XPO). The pump pulse, at a wavelength of 567 nm, has a
duration of 3 ns and a rep rate of 50 Hz. The wavelength of the pump
source is chosen to avoid damage through absorption of light by
gallium phosphide, and to obtain high enough gain of the dye. The
pump light is spatially filtered, and focused with a microscope
objective (numerical aperture, 0.55) onto the sample (focus area, 3
$\pm$ 1 $\mu$m$^2$). The pump energy on the sample was at maximum
0.32 $\mu$J/pulse. Emission light of the random laser is collected
through the microscope objective, dispersed by a spectrometer
(resolution, 1 cm$^{-1}$) and detected by an electron-multiplier
charged-coupled device (C9100-02, Hamamatsu) \footnote{The slit of
the spectrometer was set to 55 $\pm$ 15 $\mu$m. A change in the
incident angle of light caused the spectral position to shift, shown
spectra were corrected.}. The responsivity of the system was
calibrated with an Ophir power meter (PD300-3W) and a HeNe-laser.

All measurements presented in this Letter are measured at an input
energy of twice the threshold input energy, except
Fig.~\ref{fig,mode_competition_without_inset}. This threshold is
defined as the collapse of full width at half maximum (FWHM) of the
emitted spectrum \cite{Lawandy1994}. The FWHM of the spectrum far
above threshold is 13 times narrower than the FWHM of the emitted
spectrum below threshold.

To discriminate between current models, and possibly stimulate new
theoretical approaches, we want to investigate whether mode
frequencies are completely determined by the realization of disorder
inside the random laser. In contrast to the frequencies of the
modes, we expect that in all cases the peak intensities will depend
on the pump energy; as we have pulse-to-pulse fluctuations of our
pump source, we expect a different overall scale factor for our peak
intensities for every shot. We collected several spectra above
threshold at a single position on the sample. In between
measurements, the sample was translated 37~$\mu$m and back. Two of
the resulting spectra are shown in
Fig.~\ref{fig,reproducability_spikes} \footnote{Hysteresis of the
stepper motor causes the sample not to return exactly to the same
position (systematic error 0.04 $\mu$m). The red spectrum is shifted
accordingly by 0.07 nm.}. The spectral positions of the modes
reproduce, proving that mode frequencies are completely determined
by the realization of disorder, and not selected by spontaneous
emission events.
\begin{figure}
\includegraphics[width=2in]{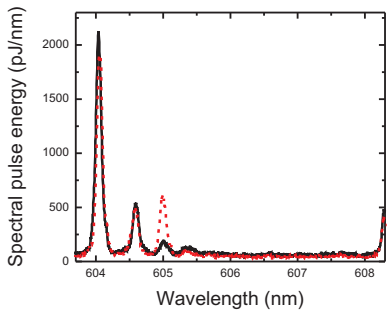}
\caption{\label{fig,reproducability_spikes}(Color) Two emission
spectra of the random laser collected at the same spot on the
sample. One spectrum (black) before, and one (red) after shifting
the sample by 37 $\mu$m and back. Apparently, the frequencies of the
modes are fully determined by the realization of the disorder.}
\end{figure}

In addition to the expected overall scale factor the peak
intensities above threshold differed individually from shot-to-shot,
as is shown in Fig.~\ref{fig,reproducability_spikes}. To investigate
the origin of this height-distribution variation, we looked at the
emission spectrum of the random laser at a single position for
different pump energies. We took 50 single shot measurements, and
used the measured pulse-to-pulse variation of the pump source to
obtain a range of pump energies. Above threshold we observed a
linear relation between the total energy of the emitted light and
the pump energy. A typical emission spectrum is shown in
Fig.~\ref{fig,mode_competition_sspectra}. We determined the energies
in the observed 4 narrow peaks in the emission spectrum (integration
of spectrum over 0.576 nm). In
Fig.~\ref{fig,mode_competition_without_inset} we plotted these
energies versus the total pulse energy of the emitted light. When
the total emitted pulse energy increases above a threshold of 50 pJ,
we observed that a single mode starts to lase (star symbol). In this
regime, the random laser can be described as a single-mode laser. We
found a decrease of the spectral width of this single mode for
increasing pulse energy in the mode, in correspondence with the
well-known Schawlow-Townes behavior \cite{Schawlow1958}. At higher
emitted pulse energies (125 to 150 pJ) two more modes of the random
laser start to lase, whereas the energy in the first mode does not
increase anymore. This leveling-off is a clear sign that the modes
are competing for the available energy, and thus are overlapping in
space. This spatial mode competition causing the height-distribution
variation is not related to the frequencies of the modes.
\begin{figure}
\includegraphics[width=2in]{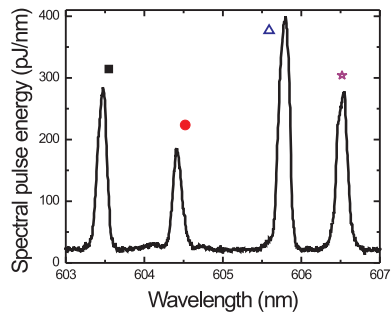}
\caption{\label{fig,mode_competition_sspectra}(Color) One of the
50 single-shot-emission spectrum of our random laser system
collected at one position of the sample. The symbols near the
peaks correspond to the pulse energy of those peaks in
Fig.~\ref{fig,mode_competition_without_inset}.}
\end{figure}

\begin{figure}
\includegraphics[width=2.5in]{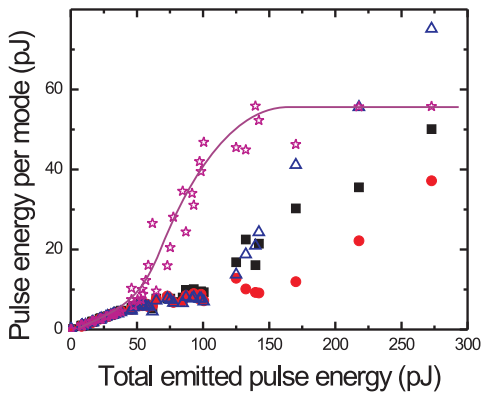}
\caption{\label{fig,mode_competition_without_inset}(Color) Pulse
energy in 4 modes [see for single-shot emission spectrum
Fig.~\ref{fig,mode_competition_sspectra}] versus the total emitted
pulse energy of the collected light for 50 single-shot spectra. The
measured errors are smaller than the symbol size. The solid line is
a guide to the eye for the star symbols. Clearly, one mode starts to
lase (star symbol), before the other modes begin to lase at higher
pump energies.}
\end{figure}

To directly observe the spatial extent of the modes inside the
random laser, we measured emission spectra while pumping at
different positions of the sample. In contrast, using speckle
correlation techniques it is possible to determine the spatial
extent of the secondary light source on the surface, originating
from the scattered light \cite{Cao2002}. In our experiment the
sample was horizontally translated with steps of 506~$\pm$~4~nm, and
for every position of the sample 50~single-shot-emission spectra
were collected. We displaced the sample 20~$\mu$m in total, which is
10~times the diameter of the pump spot. In
Fig.~\ref{fig,upcoming_spike} zoom-in of collected spectra at 5
consecutive positions are shown \footnote{Spectra of
Fig.~\ref{fig,upcoming_spike} are shifted with respect to the top
one by 0.154~nm, 0.211~nm, 0.274~nm, 0.288~nm, respectively.}.
Figure~\ref{fig,upcoming_spike} is a visualization of a typical
evolution of a peak in the emission spectrum as the position of the
sample is changed. First a peak becomes visible. Additional
displacement of the sample leads to an intensity going through a
maximum. The increase and decrease of intensity corresponds to the
overlap between the gain region and the mode. The spatial extent of
mode is defined by the difference in sample position where the mode
rises above noise level and the position where the mode falls below
the noise level (noise level is 5\%). In the specific case of
Fig.~\ref{fig,upcoming_spike} the spatial extent is
2.5~$\pm$~0.5~$\mu$m.
\begin{figure}
 \includegraphics[width=2.5in]{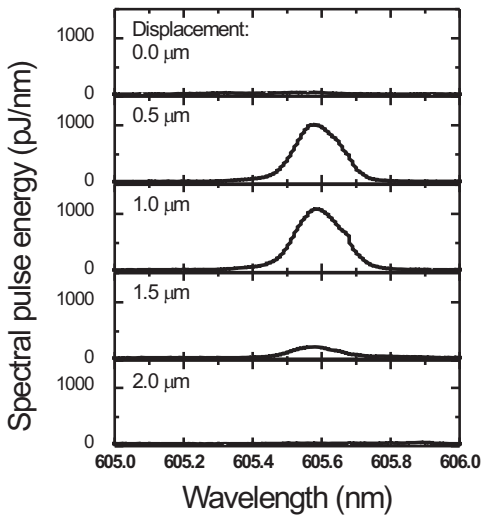}
\caption{\label{fig,upcoming_spike} Enlargement of a part of the
emission spectra where a peak is visible. Between consecutive
spectra, the sample is translated by 506~nm.}
\end{figure}

In Fig.~\ref{fig,spatial_distribution} the distribution of the
spatial extent of all the modes is plotted, together with a marker
of the diameter of the focus of the pump source on the sample.
Almost 80\% of the modes have a spatial extent smaller than the
diameter of the spot size of the focus (2~$\mu$m). The other 20\% of
the modes are larger than the diameter of the spot size of the
focus. For the LM (OM) model, we would expect all modes to have a
spatial extent much smaller (larger) than the focus diameter of the
pump spot. Apparently, our measured distribution disagrees with the
expectation of either of the two models. The selection criterium for
a mode to lase is its dwell time \cite{Molen2006a}. Our observed
limited spatial extent of lasing modes suggests that modes with long
dwell times have small spatial extent. This connection has never
been put forward before, and may be unique to three dimensional
strongly scattering media \cite{Noginov2004}.

\begin{figure}
 \includegraphics[width=3in]{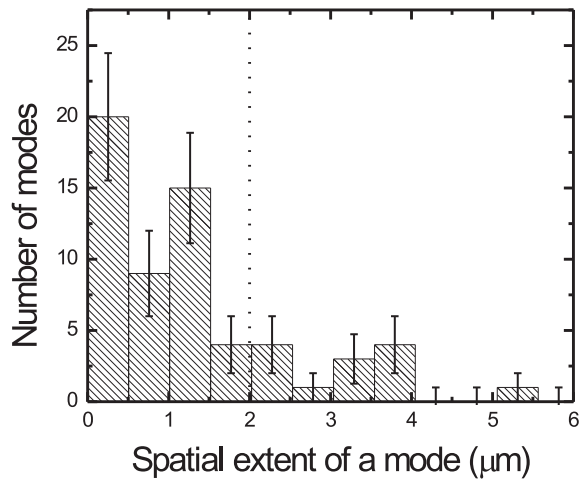}
\caption{\label{fig,spatial_distribution} Measured distribution of
the spatial size of the modes determined with a displacement
measurement. The dotted line is a marker for the diameter of the
focus of the pump source on the sample.}
\end{figure}

We will now focus on the statistics of spectral spacings of the
modes. For 11 spectra taken at widely spaced positions, we
determined the spectral spacing between two adjacent peaks. These 11
spectra are statistically independent but equivalent. The only
difference is the position on the sample where the data was
collected. Therefore we combine the 41 spectral spacings of these
11~spectra in one distribution. In Fig.~\ref{fig,level_repulsion} we
plot this distribution of the spectral mode spacing. Surprisingly,
our observations show that the statistics are determined by level
repulsion, in complete analogy to level repulsion in quantum
mechanics \cite{Beenakker1997,Patra2000}. A first experimental
indication of spectral level repulsion was shown by Cao \textit{et
al.} \cite{Cao2003b}.

To describe level repulsion quantitatively we compare our observed
statistics of the spectral mode spacings to the results of
random-matrix theory. Although random matrix theory does not
necessarily apply here, it does describe avoided crossings. The
distribution of mode spacings according to the Gaussian orthogonal
ensemble (GOE) is given by Wigner's surmise \cite{Guhr1998}
\begin{eqnarray}
P(x) = Cxe^{- \pi x^2 / (4 \Delta^2)}, \label{eqn,Wigner}
\end{eqnarray}
where $P$ is the distribution of the mode spacing $x$ of the modes,
$\Delta$ the mean mode spacing, and $C$ a scaling factor. In
Fig.~\ref{fig,level_repulsion} we plot the fit of
Eq.~(\ref{eqn,Wigner}) (black line), with a mean GOE mode spacing
$\Delta$ of 16.2~$\pm$~0.9~cm$^{-1}$. When considering all modes in
a multiple scattering system the level statistics is known to be
Poissonian (exponential) \cite{Guhr1998}. In this experiment we only
observe those modes that lase and they show level repulsion.
Apparently, the selection mechanism for random lasing causes the
level statistics to change from Poissonian statistics to level
repulsion.

\begin{figure}
 \includegraphics[width=3in]{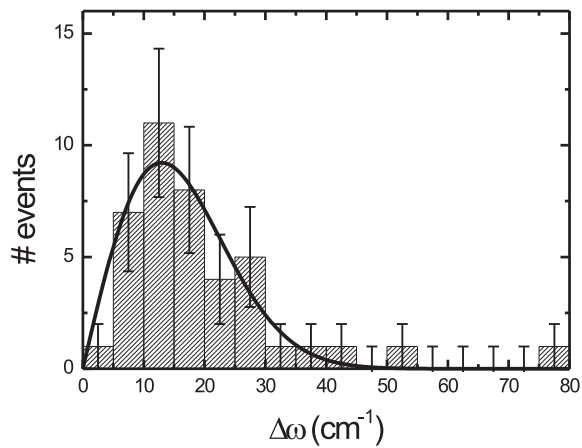}
\caption{\label{fig,level_repulsion}Measured distribution of the
spectral mode spacing for 11 emission spectra. The solid line is a
fit of Wigner's surmise~(\ref{eqn,Wigner}). The fit parameter is the
mean mode spacing $\Delta$, we find $\Delta$~=~16.2~cm$^{-1}$.}
\end{figure}

We have performed a systematic study of random lasing in porous
gallium phosphide. The frequencies of the modes in our sample are
determined by the realization of disorder. We have observed a
variation in the height distribution, and attributed this variation
to spatial mode overlap and gain competition. The distribution of
the spectral mode spacings shows clear level repulsion and can be
described with the Gaussian orthogonal ensemble. In addition the
(small) spatial extent of lasing modes was observed. We hope that
our observations stimulate more theoretical work.

The authors thank Boris Bret for preliminary measurements, and Paolo
Scalia and Sanli Faez for their help with the characterization
measurements. We acknowledge discussions with Carlo Beenakker on the
possible origins of level repulsion in random lasers. This work is
part of the research program of the 'Stichting voor Fundamenteel
Onderzoek der Materie' (FOM), which is financially supported by the
'Nederlandse Organisatie voor Wetenschappelijk Onderzoek' (NWO).

\end{document}